\documentclass[pra,superscriptaddress, showpacs, twocolumn]{revtex4}
\usepackage{newlfont}
\usepackage{amssymb}
\usepackage{amsfonts}
\usepackage{amsmath}
\usepackage{graphicx}
\usepackage{bm}
\newcommand{\ket}[1]{|#1\rangle}
\newcommand{\bra}[1]{\langle#1|}

\newcommand{\Tr}{\textrm{Tr}}

\newcommand{\kb}[2]{\ensuremath{| #1 \rangle\!\langle #2 |}}

\begin{document}

\title{Evaluable multipartite entanglement measures: are multipartite concurrences entanglement monotones?}

\author{Rafa{\l} Demkowicz-Dobrza{\'n}ski}
\affiliation{Center for Theoretical Physics, Polish Academy of Sciences, Aleja
Lotnik{\'o}w 32/44, 02-668 Warszawa, Poland}
\author{Andreas Buchleitner}
\affiliation{Max-Planck-Institut f\"ur Physik komplexer Systeme,
N\"othnitzerstr. 38, D-01187 Dresden, Germany}
\author{Marek Ku{\'s}}
\affiliation{Center for Theoretical Physics, Polish Academy of Sciences, Aleja
Lotnik{\'o}w 32/44, 02-668 Warszawa, Poland} \affiliation{Faculty of
Mathematics and Sciences, Cardinal Stefan Wyszy\'nski University, Warszawa,
Poland}
\author{Florian Mintert}
\affiliation{Max-Planck-Institut f\"ur Physik komplexer Systeme,
N\"othnitzerstr. 38, D-01187 Dresden, Germany}
\affiliation{Department of Physics,
Harvard University, Cambridge, MA 02138}

\begin{abstract}
We discuss the monotonicity of systematically constructed quantities aiming at the quantification of the
entanglement properties of multipartite quantum systems, under local operations and classical communication (LOCC).
We provide a necessary and sufficient condition for the monotonicity of generalized multipartite concurrences
which qualifies them as legitimate entanglement measures.
\end{abstract}
\maketitle

\section{Introduction}
\label{sec:intro}
Quantum correlations of composite systems play a crucial role in the rapidly
developing theory of quantum information,
since they constitute the most important resource for
applications in which quantum information processing outperforms its classical
counterpart.
A proper description and quantification of quantum entanglement,
especially for multipartite systems as the physical basis of quantum
computing devices, thus becomes an important theoretical challenge.

All quantum correlations inscribed in the simplest composite systems consisting of only
two parts are qualitatively equivalent. This means that any entangled state
can be obtained from a maximally entangled Bell-like state via local quantum
operations on each subsystem, supplemented by classical exchange of
information, concerning, {\it e.g.}, the results of various local measurements
to condition subsequent operations on the other subsystem.
The ensemble of such Local Operations with Classical Communication is summarized under the acronym LOCC.

The situation becomes much more involved when passing to multipartite systems. It is
{\it e.g.}\ possible that only some subsystems are entangled among them, being
simultaneously not correlated with the rest. But even if all subsystems are
entangled in a genuine way their correlations can be inequivalent in the sense
outlined above. For example, in a system of three qubits, both GHZ
\cite{greenberger1989} as well as W \cite{dur2000} states, arguably deserve the
label `maximally entangled'.
However, they cannot be transformed into each other
via LOCC operations \cite{dur2000}.
This mere observation clearly destroys any hope to characterize entanglement in multipartite systems with a
single scalar quantity --
what is in contrast to the bipartite case, where this is possible at least in the limit of asymptotically many states
\cite{bennett1996}.

There is a large family of quantities that characterize multipartite entanglement
\cite{mintert2005}, obtained from a generalization of the original concept of
concurrence \cite{wootters1998} for bipartite systems. Unlike most other
entanglement measures, these multipartite concurrences can be evaluated
efficiently even for mixed states \cite{mintert2005}, and they proved useful
for investigations of various problems, such as the distinction of different classes of
entanglement or, the preparation and time evolution of entanglement under decoherence \cite{carvalho2005,mintert2005a}.

The construction of multipartite concurrences is based on the observation that the concurrence of
a pure state $\ket\Psi$ of a bipartite system with Hilbert space
$\mathcal{H}=\mathcal{H}_1\otimes\mathcal{H}_2$ can be expressed by the
expectation value
\begin{equation}
\label{eq:concurrence}
C(\ket{\psi})=2\sqrt{\bra{\psi} \otimes \bra{\psi}\mathcal{A}\ket{\psi}\otimes \ket{\psi}},
\end{equation}
where $\mathcal{A}$ is the projection operator from $\mathcal{H}\otimes{\cal H}$ onto its
locally antisymmetrized part
$\mathcal{H}_1\wedge\mathcal{H}_1\otimes\mathcal{H}_2\wedge\mathcal{H}_2$.
This expression is extended to mixed states $\varrho$ by the {\it convex roof}
\begin{equation}
\label{concurrencemixed}
C(\rho)=\mathrm{min}_{\psi_i}\sum_iC(\ket{\psi_i})\ ,
\end{equation}
where the minimum of the {\it average concurrence} $\sum_iC(\ket{\psi_i})$ is
taken over all pure state decompositions $\rho=\sum_i\kb{\psi_i}{\psi_i}$ with sub-normalized pure states $\ket{\psi_i}$ .

To generalize the above concept for an $N$-partite system on a Hilbert space
$\mathcal{H}=\mathcal{H}_1\otimes\cdots\otimes\mathcal{H}_N$, we consider
projections $P^{(i)}_-$ and $P^{(i)}_+$ on the antisymmetric and symmetric
subspaces -- $\mathcal{H}_i\wedge\mathcal{H}_i$ and
$\mathcal{H}_i\odot\mathcal{H}_i$ -- of $\mathcal{H}_i\otimes\mathcal{H}_i$,
and form a linear operator acting on $\cal{H}\otimes\cal{H}$ as a linear
combination of their tensor products
\begin{equation}
\label{A}
\mathcal{A}=\sum_{s_1,\dots,s_N} p_{s_1,\dots,s_N}\
P_{s_{1}}\otimes P_{s_{2}}\otimes \dots \otimes P_{s_{N}},
\end{equation}
where $s_i\in\{-,+\}$ (for simplicity of notation, we dispose of superscripts
indicating subsystems, since the order in which the individual projectors enter
Eq.~(\ref{A}) corresponds to their numbering).
For Eq.~(\ref{eq:concurrence}) to make sense, we set $p_{s_1,\dots,s_N}\geq 0$ --
this renders the operator $\mathcal{A}$ positive semi-definite.

Each operator $\mathcal{A}$ can be used
to define a generalized concurrence for a pure state $\ket{\psi}\in\mathcal{H}$
exactly as in the bipartite case,
$C_\mathcal{A}(\ket{\psi})=2\sqrt{\bra{\psi}\otimes\bra{\psi}\mathcal{A}\ket{\psi}\otimes\ket{\psi}}$,
and we use a notation which makes the dependence on $\mathcal{A}$ explicit.
Since terms with an odd number of projectors on antisymmetric spaces give a vanishing contribution,
we can set the corresponding coefficients $p_{s_1,\dots,s_N}$ equal to zero, without loss of generality.
Analogously to the bipartite case, the extension to mixed
states is achieved by the convex roof (\ref{concurrencemixed})
\cite{mintert2005}.

By construction, $C_\mathcal{A}(\ket{\psi})$ is invariant under local unitary
transformations, and it allows to discriminate between qualitatively different
entangled states \cite{mintert2005}. Though, {\it a priori} nothing is said
about its behavior under LOCC.
Monotonic decrease under LOCC is {\it the} crucial property of an entanglement measure that allows
not merely to quantify correlations of a given state but also to decide whether
one state can be obtained from another one by LOCC \cite{vidal2000}.
In particular, if
there exist two different concurrences $C_{\mathcal{A}_1}$ and
$C_{\mathcal{A}_2}$, proved to be entanglement monotones, that establish different orders of states $\ket{\phi}$ and
$\ket{\psi}$, {\it e.g.}
$C_{\mathcal{A}_1}(\ket{\phi})<C_{\mathcal{A}_1}(\ket{\psi})$, whereas
$C_{\mathcal{A}_2}(\ket{\phi})>C_{\mathcal{A}_2}(\ket{\psi})$) then no conversion between
$\ket{\phi}$ and $\ket{\psi}$ via LOCC is possible.

\section{Entanglement monotones}
\label{sec:monotonicity} In general, it can be rather involved to prove
monotonicity of a multi-partite quantity by considering the action of the most
general LOCC operation on it. Though, there is a systematic approach to find
necessary and sufficient conditions for a quantity to be an entanglement
monotone \cite{vidal2000}. The problem of monotonicity was further simplified
\cite{horodecki2004} by proving that a convex
function $f$ is an entanglement monotone if and only if
\begin{itemize}
\item[-]{$f$ is invariant under local unitaries, and}
\item[-]{satisfies the so-called `FLAGS condition':
\begin{equation}
\label{eq:flags} f\left(\sum_i p_i \rho^{(i)} \otimes \ket{\eta_{i}}\bra{\eta_{i}}\right)=
\sum_i p_i f\left(\rho^{(i)}\right),
\end{equation}
for all multipartite states $\rho^{(i)}$, and all mutually orthogonal
local states $\ket{\eta_{i}}$ -- called flags -- attached to the same, arbitrarily chosen subsystem.}
\end{itemize}

The FLAGS condition is equivalent to a simpler one with only two terms in the
sum, {\it i.e.}
\begin{widetext}
\begin{equation}
\label{eq:flags2}
f\left(p_1 \rho^{(1)} \otimes \ket{\eta_1} \bra{\eta_1}+p_2\rho^{(2)}
\otimes \ket{\eta_2}\bra{\eta_2} \right)= p_1 f\left(\rho^{(1)}\right)+p_2
f\left(\rho^{(2)}\right).
\end{equation}
\emph{Proof.} 
Let $\tilde{\rho}=\sum_i p_i \rho^{(i)} \otimes \ket{\eta_{i}}\bra{\eta_i}$. Consider 
the state $\tilde{\rho}\otimes \ket{\eta_0}\bra{\eta_0}$ (flag $\ket{\eta_0}$ is added to the same subsystem as flags $\ket{\eta_i}$). 
From Eq.~(\ref{eq:flags2}) it follows in particular that $f(\tilde{\rho} \otimes \ket{\eta_0}\bra{\eta_0}) = f(\tilde{\rho})$.
Moreover, the state $\tilde{\rho} \otimes \ket{\eta_0}\bra{\eta_0}=\sum_i p_i \rho^{(i)} \otimes \ket{\eta_{i}}\bra{\eta_i}\otimes \ket{\eta_0}\bra{\eta_0}$
can be transformed by local unitaries (local with respect to subsystems) into the state: 
$p_1 \rho^{(1)} \otimes \ket{\eta_1}\bra{\eta_1}\otimes \ket{\eta_1}\bra{\eta_1} +
\left[\sum_{i\ge 2} p_i \rho^{(i)} \otimes \ket{\eta_i}\bra{\eta_i}\right] \otimes
\ket{\eta_2}\bra{\eta_2}$. Therefore, thanks to the invariance of $f$ under local unitaries we 
can write:
\begin{equation}
f\left(\sum_i p_i \rho^{(i)} \otimes\ket{\eta_i}\bra{\eta_i}\right)=
f\left(p_1 \rho^{(1)} \otimes \ket{\eta_1}\bra{\eta_1}\otimes \ket{\eta_1}\bra{\eta_1} +
\left[\sum_{i\ge 2} p_i \rho^{(i)} \otimes \ket{\eta_i}\bra{\eta_i}\right] \otimes
\ket{\eta_2}\bra{\eta_2}\right).
\end{equation}
Applying Eq.~(\ref{eq:flags2}) we obtain:
\begin{equation}
f\left(\sum_i p_i \rho^{(i)} \otimes \ket{\eta_i}\bra{\eta_i}\right)= p_1
f\left(\rho^{(1)}\right) + f\left(\sum_{i \ge 2} p_i \rho^{(i)} \otimes
\ket{\eta_i}\bra{\eta_i}\right).
\end{equation}
By induction, this implies  Eq.~(\ref{eq:flags}).
\end{widetext}

Adding local flags to a subsystem does not change the number of parties, but
increases the dimension of one subsystem. Consequently, the function $f$ needs
to be defined for arbitrary (finite) dimensional systems, as concurrence in
Eq.~(\ref{eq:concurrence}) and most of the commonly used entanglement monotones
are. Quantities that are defined solely for systems of specific dimensions
cannot be characterized with the above criterion, and emerging consequences are
addressed in Sec.~\ref{sec:controverse}.

\section{Simple criterion for monotonicity}

With a few more assumptions on a potential monotone, one can reduce
Eq.~(\ref{eq:flags}) to an even simpler criterion, which eventually will allow
us to investigate the monotonicity properties of the multipartite concurrences in question
here.

A function $C$ that
\begin{itemize}
\item[-]is real, nonnegative and invariant under local unitaries,
\item[-]satisfies {$C(a\ket{\psi})=|a|^2 C(\ket{\psi})$}, and
\item[-]is defined for mixed states as a convex roof, c.f.\ Eq.~(\ref{concurrencemixed}),
\end{itemize}
is an entanglement monotone if and only if
\begin{equation}
\label{eq:condition1}
 {C}(a \ket{\psi} \otimes \ket{\eta_{1}}+b\ket{\phi} \otimes\ket{\eta_{2}}) \geq
 |a|^2 {C}(\ket{\psi}) + |b|^2 {C}(\ket{\phi})\ ,
\end{equation}
with equality for $a=0$ or $b=0$,
where $\ket{\phi}$ and $\ket{\psi}$ are arbitrary multipartite pure states,
and $\ket{\eta_{1}}$, $\ket{\eta_{2}}$ are local orthogonal flags. This criterion clearly
applies -- though not exclusively -- to the presently considered multipartite concurrences.

\textit{Proof of sufficiency.} In order to show that $C$ is an entanglement
monotone if it satisfies Eq.~(\ref{eq:condition1}), let us take two arbitrary
multipartite states $\rho^{(1)}$, $\rho^{(2)}$ with respective optimal
decompositions
$\rho^{(i)}=\sum_{j}|\psi_{j}^{(i)}\rangle\langle\psi_{j}^{(i)}|$, ($i=1,2$),
which minimize the average concurrence.
We show that the FLAGS condition, Eq.~(\ref{eq:flags2}),
\begin{equation}
\label{eq:proof}
C(\tilde{\rho})= p_1 C(\rho^{(1)}) + p_2 C(\rho^{(2)})
\end{equation}
holds, where
$\tilde{\rho}=p_1\rho^{(1)}\otimes \ket{\eta_{1}}\bra{\eta_{1}} + \rho^{(2)}\otimes\ket{\eta_{2}}\bra{\eta_{2}}$,
and $\ket{\eta_{1}}$, $\ket{\eta_{2}}$ are the local flags.
The decomposition consisting of the vectors:
\begin{equation}
\label{eq:decomp}
\sqrt{p_1}\ket{\psi_{j}^{(1)}}\otimes \ket{\eta_{1}},\ \mbox{and}\
\sqrt{p_2}\ket{\psi_{j}^{(2)}} \otimes \ket{\eta_{2}}
\end{equation}
 is a valid decomposition of $\tilde{\rho}$,
 and hence due to Eq.~(\ref{concurrencemixed}) we have the inequality
\begin{equation}
\label{eq:ineqtmp}
C(\tilde{\rho}) \leq  p_1 C(\rho^{(1)} \otimes \ket{\eta_{1}}\bra{\eta_{1}}) +
p_2 C(\rho^{(2)}\otimes \ket{\eta_{2}}\bra{\eta_{2}}).
\end{equation}
Notice that thanks to assumed equality in Eq.~(\ref{eq:condition1}), whenever $a=0$ or $b=0$, 
we have $C(\ket{\psi}\otimes \ket{\eta})=C(\ket{\psi})$, 
and consequently $C(\rho \otimes \ket{\eta}\bra{\eta})=C(\rho)$. Inequality (\ref{eq:ineqtmp})
thus reads:
\begin{equation}
C(\tilde{\rho}) \leq  p_1 C(\rho^{(1)}) +p_2 C(\rho^{(2)}).
\end{equation}
In order to prove equality, and thus establish Eq.~(\ref{eq:proof}),
it is enough to show that using coherent superpositions of the form
\begin{equation}
\label{eq:super}
a\ket{\psi}\otimes \ket{\eta_{1}}+b\ket{\phi}\otimes \ket{\eta_{2}}\ 
\end{equation}
in the decomposition of $\tilde{\rho}$, one can only increase the average
concurrence. This is, however, exactly the claim of Eq.~(\ref{eq:condition1}).
Consequently, if a decomposition of $\tilde{\rho}$ contains states that are 
superpositions of terms with different local flags, then it is preferable for each of them to
take two states forming a superposition as two separate vectors in the
decomposition, and thus lower the average concurrence. Notice that this always
leads us to a valid decomposition, since $\tilde{\rho}$ does not contain any coherences 
between states with different flags $\ket{\eta_1}$, $\ket{\eta_2}$.
The optimal decomposition for
$\tilde{\rho}$ is thus composed of vectors forming optimal decompositions of
$p_1\rho^{(1)} \otimes \ket{\eta_{1}} \bra{\eta_{1}}$ and
$p_2\rho^{(2)} \otimes \ket{\eta_{2}}\bra{\eta_{2}}$
and 
Eq.~(\ref{eq:proof}) is satisfied.

\textit{Proof of necessity.} To show that any $C$ necessarily needs to satisfy
Eq.~(\ref{eq:condition1}) in order to be an entanglement monotone,
we prove ad absurdum. Let us assume that inequality
Eq.~(\ref{eq:condition1}) is violated for $a\ket{\psi}$, and $b\ket{\phi}$.
Consequently, introducing the state 
$\ket{\xi_{1}}=(a \ket{\psi}\otimes \ket{\eta_{1}} + b\ket{\phi} \otimes
\ket{\eta_2})/\sqrt{2}$
we have
\begin{equation}
\label{eq:better1}
C(\ket{\xi_1}) <  C(a\ket{\psi})/2 + C(b\ket{\phi})/2\, .
\end{equation}
The state 
$\ket{\xi_2}=(a \ket{\psi}\otimes \ket{\eta_{1}} - b\ket{\phi} \otimes \ket{\eta_2})/\sqrt{2}$
can be obtained from $\ket{\xi_1}$ via local unitaries, thus $C(\ket{\xi_1})=C(\ket{\xi_2})$, which implies:
\begin{equation}
\label{eq:better2}
C(\ket{\xi_2}) < C(a\ket{\psi})/2 + C(b\ket{\phi})/2.
\end{equation}
If the states $a\ket{\psi}\otimes \ket{\eta_{1}}$ and $b\ket{\phi}\otimes \ket{\eta_{2}}$
provided an optimal decomposition for
$\tilde{\rho}=|a|^2\ket{\psi}\bra{\psi}\otimes\ket{\eta_{1}}\bra{\eta_{1}}+
|b|^2 \ket{\phi} \bra{\phi}\otimes\ket{\eta_{2}}\bra{\eta_{2}}$,
the criterion of Eq.~(\ref{eq:flags2}) would be satisfied.
However, due to Eqs.(\ref{eq:better1},\ref{eq:better2})
replacing $a\ket{\psi}\otimes \ket{\eta_1}$ and $b\ket{\phi} \otimes \ket{\eta_2}$ 
with states $\ket{\xi_1}$, $\ket{\xi_2}$ yields a valid decomposition of $\tilde{\rho}$ with a smaller concurrence.
Therefore:
\begin{equation}
C(\tilde{\rho}) < |a|^2 C(\ket{\psi}\bra{\psi}) + |b|^2 C(\ket{\phi}\bra{\phi})\ ,
\end{equation}
which is in contradiction with the FLAGS condition (\ref{eq:flags2}), and
consequently $C$ is not an entanglement monotone. 
$\blacksquare$

\section{Monotonicity of multipartite concurrence}
\label{sec:monconcurrence}
Now we are ready to address the question of monotonicity of the concurrences
$C_{\mathcal A}$, defined in Eqs.~(\ref{eq:concurrence}), and (\ref{A}),
through the non-negative parameters $p_{s_{1},\hdots,s_{N}}$.
We want concurrence to vanish on completely separable states
(states that are separable with respect to any partition).
The only term in Eq.~(\ref{A}) that yields a non-zero contribution
for such states is the term $P_{+}\otimes\hdots\otimes P_{+}$.
Therefore, we set its prefactor $p_{+,\dots,+}$ equal to zero.
Consequently, the concurrence depends on $2^{N-1}-1$ parameters $p_{s_1,\dots,
s_N}$. Let us express $C_{{\cal A}}$ explicitly in terms of partial traces
\begin{equation}
\label{eq:multiconc}
C_\mathcal{A}(\ket{\psi})=2^{1-N/2}\sqrt{
\sum_{\mathbf{V}}\alpha_\mathbf{V}
\Tr\left(\Tr_\mathbf{V}(\ket{\psi}\bra{\psi})^2 \right)}\ ,
\end{equation}
where the summation is performed over all subsets $\mathbf{V}$ of
$\mathbf{N}=\{1,\dots,N\}$, and the prefactors $\alpha_{\mathbf{V}}$ are given
in terms of the $p_{s_1,\dots,s_N}$ via
\begin{equation}\label{alpha-p}
\alpha_\mathbf{V}=\sum\limits_{s_1,\dots,s_N} p_{s_1,\dots,s_N}
\prod\limits_{i\in \mathbf{V}} s_i\ .
\end{equation}
From the above formula it follows that $\alpha_{\mathbf{V}}=\alpha_{\bar{\mathbf{V}}}$, where $\bar{\mathbf{V}}$ denotes
the complement set to $\mathbf{V}$. Furthermore, notice that $\sum_\mathbf{V} \alpha_\mathbf{V}=2^N p_{+,\dots,+}$, which is zero according to
what was said before. Consequently, we have $2^{N-1}-1$ independent parameters $\alpha_\mathbf{V}$ describing the concurrence
$C_\mathcal{A}$.
The parameterizations via $p_{s_1,\dots,s_N}$ and $\alpha_\mathbf{V}$ are equivalent, as one can invert the relation
(\ref{alpha-p}):
\begin{equation}
\label{eq:pis} p_{s_1,\dots,s_N}=\frac{1}{2^N}\sum_{\mathbf{V}}
\alpha_\mathbf{V} \prod_{i\in \mathbf{V}}s_i\ .
\end{equation}
When parameterizing $C_\mathcal{A}$ using $\alpha_\mathbf{V}$,
one has to keep in mind the positivity condition on $p_{s_1,\dots,s_N}$
which requires that $\sum_{\mathbf{V}} \alpha_\mathbf{V} \prod_{i\in \mathbf{V}}s_i \geq 0$, for every choice of $s_1,\dots,s_N$.

Without loosing any relevant information on entanglement, we can normalize the concurrence.
We adopt here the convention
\begin{equation}
\label{eq:norm}
\sum_{s_1,\dots,s_N} p_{s_1,\dots,s_N}=2^{N-1}-1\ ,
\end{equation}
so that the maximum value of concurrence for bipartite qubit states equals unity.
In terms of the $\alpha_\mathbf{V}$, this constraint amounts to setting:
$\alpha_{\mathbf{\varnothing}}=\alpha_{\mathbf{N}}=2^{N-1}-1$, or, equivalently,
\begin{equation}
\label{eq:sumalpha}
\sum_{\mathbf{V}\neq \mathbf{\varnothing},\mathbf{N}} \alpha_\mathbf{V}=-(\alpha_\varnothing + \alpha_\mathbf{N})=-(2^N-2)\ .
\end{equation}
In particular, the choice $\alpha_\mathbf{V}=-1$ for each
$\mathbf{V}\neq\mathbf{\varnothing},\mathbf{N}$ leads to the symmetric concurrence $C_S$,
used in \cite{carvalho2005,mintert2005a} to investigate the dynamics of decoherence in
various systems. Thanks to the chosen normalization it enjoys the property
$C_{S}\left(\ket{\psi}_{1,\dots,N}\right)= C_S\left(\ket{\psi}_{1,\dots,N} \otimes \ket{\phi}_{N+1}\right)$:
When an $(N+1)$--st party is added to the original $N$ parties as a product state,
the symmetric concurrence remains unchanged (Do not confuse adding an $(N+1)$--st
party with adding a local flag to one of the parties).

Thanks to the simple parameterization of $C_{\mathcal A}$, it is possible to
give explicit conditions for monotonicity in terms of the parameters
$\alpha_\mathbf{V}$.

\subsection{Sufficient condition for monotonicity}

$C_\mathcal{A}$ is an entanglement monotone if
\begin{equation}
\label{theorem}
\alpha_\mathbf{V} \leq 0\ ,\ \textrm{ for all } \mathbf{V}\neq \mathbf{\varnothing},\mathbf{N}.
\end{equation}

\textit{Proof:} In order to shorten the following formulae we shall adopt the
notation $\Tr_{\mathbf{V}}(\ket{\psi}\bra{\psi})=\rho^\psi_{\bar{\mathbf{V}}}$.
Making use of the explicit parameterization of $C_{\mathcal A}$ given in
Eq.~(\ref{eq:multiconc}), the inequality (\ref{eq:condition1}) to be proven
reads
\begin{equation}
\begin{array}{l}
\label{eq:multiineq}
\sqrt{\sum\limits_{\mathbf{V}} \alpha_\mathbf{V}
\Tr\left(\left[\Tr_\mathbf{V}\ket{\Xi}\bra{\Xi}\right]^2  \right)} \geq \\
|a|^2\sqrt{\sum\limits_{\mathbf{V}}\alpha_\mathbf{V}\Tr(\rho^{\psi}_{\bar{\mathbf{V}}})^2} +
|b|^2 \sqrt{\sum\limits_{\mathbf{V}}\alpha_\mathbf{V} \Tr(\rho^{\phi}_{\bar{\mathbf{V}}})^2}\ ,
\end{array}
\end{equation}
where $\ket{\Xi}=a\ket{\psi}\otimes \ket{\eta_{1}} +b\ket{\phi}\otimes\ket{\eta_{2}}$,
and, without loss of generality, we can assume that the local flags are added to the first subsystem.
Moreover, we have
\begin{equation}
\Tr\left(\left[\Tr_\mathbf{V}(\ket{\Xi}\bra{\Xi}\right]^2\right) = |a|^4 \Tr(
\rho^{\psi}_{\bar{\mathbf{V}}})^2+ |b|^4 \Tr(
\rho^{\phi}_{\bar{\mathbf{V}}})^2+2|a|^2|b|^2 \Upsilon_\mathbf{V},
\end{equation}
where
$\Upsilon_\mathbf{V} = \Tr\rho^\psi_{\bar{\mathbf{V}}} \rho^\phi_{\bar{\mathbf{V}}}$
if
$\mathbf{V}\ni 1$, and
$\Upsilon_\mathbf{V} =
\Tr\rho^\psi_{\mathbf{V}}\rho^\phi_{\mathbf{V}}$ otherwise.
Squaring inequality
(\ref{eq:multiineq}) we arrive at:
\begin{equation}
\label{eq:multiineq2}
\sum_{\mathbf{V}}\alpha_\mathbf{V} \Upsilon_\mathbf{V}\geq
\sqrt{\sum_{\mathbf{V}} \alpha_\mathbf{V}\Tr(\rho^{\psi}_{\bar{\mathbf{V}}})^{2}}
\sqrt{\sum_{\mathbf{V}} \alpha_\mathbf{V}\Tr(\rho^{\phi}_{\bar{\mathbf{V}}})^{2}}\ .
\end{equation}
As a direct consequence of the Cauchy-Schwarz inequality,
we have
\begin{equation}
\label{eq:cauchy} \Upsilon_\mathbf{V} \leq \frac{1}{2} \left[
\Tr(\rho^{\psi}_{\bar{\mathbf{V}}})^2+
\Tr(\rho^{\phi}_{\bar{\mathbf{V}}})^2 \right]\ ,
\end{equation}
and equality holds for the cases $\mathbf{V}=\mathbf{\varnothing},\mathbf{N}$
where $\Upsilon_{\mathbf{\varnothing}}=\Upsilon_{\mathbf{N}}=1$. Now, since
$\alpha_{\mathbf{V}}\leq 0$ holds by assumption for
$\mathbf{V}\neq\mathbf{\varnothing},\mathbf{N}$, it is legitimate to use
Eq.~(\ref{eq:cauchy}), and substitute for $\Upsilon_{\mathbf{V}}$ in
Eq.~(\ref{eq:multiineq2}) the right-hand-side of (\ref{eq:cauchy}).
This will lower the left hand side of Eq.~(\ref{eq:multiineq2}).
Thus, if we are able to prove the resulting inequality, the original one
(\ref{eq:multiineq2}) will follow.
After some further algebra we arrive at:
\begin{equation}
\frac{1}{4}\left[\sum_{\mathbf{V}} \alpha_\mathbf{V} \left[
\Tr(\rho^{\psi}_{\bar{\mathbf{V}}})^{2}-
\Tr(\rho^{\phi}_{\bar{\mathbf{V}}})^{2} \right]\right]^2 \geq 0\ ,
\end{equation}
which is obviously true and concludes the reasoning. $\blacksquare$

\subsection{Necessary condition for monotonicity}

One may now wonder whether the above sufficient criterion is also necessary for
a multipartite concurrence to be an entanglement monotone. In this subsection
we show that this is indeed the case for bi--, and tri--partite systems,
{\it i.e.} for $N=2$ and $N=3$. For the four-partite case, however, (and as a
consequence also for cases involving more than four parties) we have at least
numerical evidence that there are concurrences with some
$\alpha_\mathbf{V} >0$, which nevertheless are still entanglement monotones.

\subsubsection{Bipartite systems}

The concurrence of bipartite systems in terms of parameters $\alpha_\mathbf{V}$ reads:
\begin{equation}
C(\ket{\psi})=\sqrt{2\alpha_\varnothing+2\alpha_{\{1\}} \Tr(\rho^{\psi}_{\{1\}})^{2}}\ ,
\end{equation}
and from the discussion at the beginning of Sec.~\ref{sec:monconcurrence} it follows that the parameters
are determined uniquely: $\alpha_\varnothing=1$, $\alpha_{\{1\}}=-1$.
Indeed, $\alpha_{\{1\}}$ is negative (from
a more general perspective, $\alpha_{\{1\}}$ has to be negative to make $C$ a concave function
of the one-party reduced density matrix, which is a necessary and sufficient condition for bipartite
quantities constructed via the convex-roof to be entanglement monotones \cite{vidal2000}).

\subsubsection{Tripartite systems}

In the case of tripartite systems, we have three relevant parameters
$\alpha_\mathbf{V}$ -- $\alpha_{\{1\}}, \alpha_{\{2\}}, \alpha_{\{3\}}$ --
which due to normalization (Eq.~(\ref{eq:sumalpha})) sum up to $-3$.
Let us assume that one of them is positive. Without loss of generality we can
take $\alpha_{\{1\}}>0$. As we will show, in this case concurrence is not an
entanglement monotone and, consequently, Eq.~(\ref{theorem}) is not only a
necessary, but also a sufficient criterion. For this purpose it is sufficient
to show that Eq.~(\ref{eq:condition1}) is violated for some states. Let us take
the two tripartite states
$\ket{\psi}=\ket{0}\otimes \ket{\Phi^+}$ and
$\ket{\phi}=\ket{0}\otimes\ket{\Phi^-}$,
with equal weights $a=b=1/\sqrt{2}$, where the state of the first subsystem is the same in both
cases, and the remaining two subsystems are in orthogonal maximally entangled
states,
$\ket{\Phi^{\pm}}=(\ket{0}\otimes\ket{0}\pm\ket{1}\otimes\ket{1})/\sqrt{2}$.
For these states one has
$\Tr(\rho^{\psi}_\mathbf{V})^{2}= \Tr(\rho^{\phi}_\mathbf{V})^{2}$ for any
$\mathbf{V}$, and additionally $\rho^\psi_\mathbf{V}=\rho^\phi_\mathbf{V}$ for
$\mathbf{V}=\{1\}, \{2\}, \{3\}, {\{1,2}\}$, and $\{1,3\}$. Consequently, after adding
local flags to subsystem 1, Eq.~(\ref{eq:multiineq}) reduces to
\begin{equation}
\label{eq:final}
 \sqrt{\sum_{\mathbf{V}}\frac{\alpha_{\mathbf{V}}}{2} \left(
 \Tr(\rho^{\psi}_{\bar{\mathbf{V}}})^{2}+\Upsilon_\mathbf{V}\right)}\geq
 \sqrt{\sum\limits_{\mathbf{V}}
 \alpha_\mathbf{V}\Tr(\rho^{\psi}_{\bar{\mathbf{V}}})^{2}}\ ,
 \end{equation}
 what, after cancellation of several terms reduces to
 \begin{equation}
\sum_{\mathbf{V}=\{1\},\{2,3\}}\frac{\alpha_{\mathbf{V}}}{2}
\left(\Tr(\rho^{\psi}_{\bar{\mathbf{V}}})^{2}-\Upsilon_{{\mathbf{V}}}\right)\leq 0\ .
 \end{equation}
Now, one easily verifies that $\Upsilon_{\{1\}}=\Upsilon_{\{2,3\}}=0$, what
leads to the conclusion that $\alpha_{\{1\}}=\alpha_{\{2,3\}}\le 0$. This,
however, is in contradiction with the assumption $\alpha_{\{1\}}>0$ made above.
Thus, no tri-partite $C_{\mathcal{A}}$ with some positive prefactor
$\alpha_{\mathbf{V}}$ can be monotonously decreasing under LOCC. And
consequently, the criterion (\ref{theorem}) is indeed also necessary.

\subsubsection{Four-partite and larger systems}
For four and more parties, the condition that one of the $\alpha_\mathbf{V}$ is
positive is not enough to show that the corresponding concurrence is not an
entanglement monotone. The counterexample valid for $N=3$ can be only applied
when that subset $\mathbf{V}$ for which $\alpha_\mathbf{V}>0$ contains only one
element -- then everything can be done analogously as in the $N=3$ case, except
that one takes
$\ket{\Phi^\pm}=(\ket{0}\otimes \dots \otimes \ket{0}
\pm \ket{1}\otimes\dots\otimes \ket{1})/\sqrt{2}$.

The idea behind this reasoning is simply to find states $\ket{\psi}$, $\ket{\phi}$ for which
$\Tr(\rho^{\psi}_\mathbf{V})^2=\Tr(\rho^{\phi}_\mathbf{V})^2$, for each
$\mathbf{V}$. This assures that, on the right hand side of
Eq.~(\ref{eq:multiineq}), both square roots are equal. Furthermore we need a
situation in which for every $\mathbf{V}$, except the one for which we assumed
$\alpha_\mathbf{V}>0$, $\Upsilon_\mathbf{V}=\Tr(\rho^{\phi}_{\bar{\mathbf{V}}})^2$. As
a result all the terms cancel out except for the one with positive
$\alpha_\mathbf{V}$, for which we need to have $\Upsilon_\mathbf{V} <
\Tr(\rho^{\phi}_{\bar{\mathbf{V}}})^2$ (in the $N=3$ case $\Upsilon_\mathbf{V}$ was zero).
If this conditions are fulfilled, we arrive at a violation of
Eq.~(\ref{eq:final}).

Hence, we conclude that, for an arbitrary number of parties, if even one coefficient
$\alpha_\mathbf{V}$ is positive for a single-element subset $\mathbf{V}$,
the corresponding concurrence is not an entanglement monotone. (Remember the
assumption that $\alpha_\mathbf{V}=\alpha_{\bar{\mathbf{V}}}$, so if
$\alpha_\mathbf{V}>0$ for some one-element subset, it is also positive for the
complementary $(N-1)$-element subset)

If, however, $\alpha_\mathbf{V}$ is positive only for subsets containing more than one
element (but less than $N-1$), then the above strategy of violating
inequality (\ref{eq:condition1}) will not work, and in principle the concurrence may be an entanglement monotone.

In particular, we have numerically checked the case $N=4$.
We set equal all coefficients $\alpha_\mathbf{V}=\kappa_{1}$
for one-element or three-element subsets $\mathbf{V}$,
and $\alpha_\mathbf{V}=\kappa_{2}$ whenever $\mathbf{V}$ is a two-element subset.
Due to normalization (\ref{eq:sumalpha}) the coefficients $\kappa_1$ and $\kappa_2$ are related: $8\kappa_1+6\kappa_2=-14$.
The legitimate range of $\kappa_1$ which is compatible with positivity of $\mathcal{A}$ is $\kappa_1 \in [-7,0 ]$ 
(and correspondingly $\kappa_2$: $\kappa_2 \in [-7/3,7 ]$).
Assuming that the  states $\ket{\phi}$, $\ket{\psi}$ are arbitrary $4$-qubit states,
we found that the inequality (\ref{eq:condition1}) is violated
if and only if  $\kappa_1 <-2.8$ (which corresponds to $\kappa_2 >1.4$).
That means that for these values of $\kappa_i$
the concurrence is surely not an entanglement monotone,
while for $\kappa_1 \geq -2.8$ (which corresponds to $\kappa_2 \leq 1.4$ -- hence, also positive $\alpha_V$ are allowed) the concurrence
is likely to be monotonous.
To be absolutely sure of that we would have to investigate higher dimensional four-partite states
$\ket{\psi}$, $\ket{\phi}$, and not only qubits as we did here.

\section{Example: tripartite monotones}
\label{sec:ex}

\begin{figure}[t]
\includegraphics[scale=0.7]{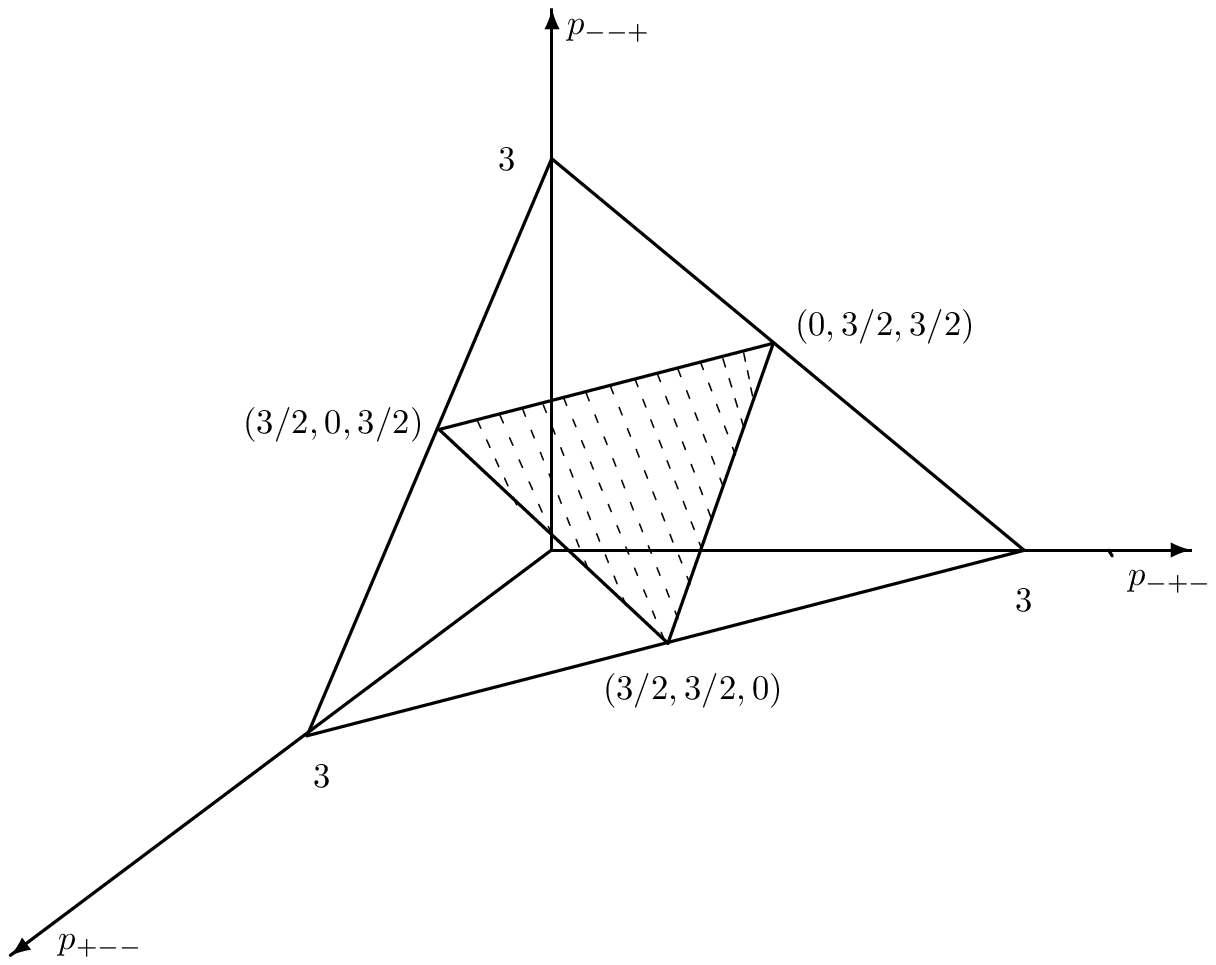}
\caption{Admissible parameters for monotonous generalized tri-partite concurrences are confined to the dashed region.}
\label{fig1}
\end{figure}

As an example, we construct all monotonous concurrence-like quantities for
tri-partite systems. From the general prescription given in Sec.~\ref{sec:intro} it is
clear that the most general form of the projection operator $\mathcal{A}$
defining the generalized pure-state concurrence reads
\begin{widetext}
\begin{equation}\label{eq:tripart-gen}
\mathcal{A}=p_{+--}P_+^{(1)}\otimes P_-^{(2)}\otimes P_-^{(3)} +
p_{-+-}P_-^{(1)}\otimes P_+^{(2)}\otimes P_-^{(3)} +
p_{--+}P_-^{(1)}\otimes P_-^{(2)}\otimes P_+^{(3)},
\end{equation}
and we assume $p_{+--}\ge 0$, $p_{-+-}\ge 0$, and $p_{--+}\ge 0$, to assure
positive-definiteness of $\mathcal{A}$. The normalization condition (Eq.~(\ref{eq:norm})) yields:
$p_{+--}+p_{-+-}+p_{--+}=3$.
From Eq.~(\ref{alpha-p}) we infer three independent conditions that guarantee
non-positivity of $\alpha_{\mathbf{V}}$:
\begin{equation}\label{eq:conditions}
p_{+--}+p_{-+-}\ge p_{--+}, \quad p_{-+-}+p_{--+}\ge p_{+--}, \quad p_{+--}+p_{+--}\ge
p_{-+-}.
\end{equation}
\end{widetext}
The resulting parameter region which is spanned by admissible values of the coefficients is depicted in
Fig.~\ref{fig1}.

\section{
Some controversies over the $P_- \otimes P_- \otimes P_- \otimes P_-$ example}
\label{sec:controverse}

As mentioned earlier, the presently used tools require a quantity that is
defined in arbitrary dimensions. And, as the following example illustrates,
there are quantities that are monotonously decreasing with respect to LOCC
operations that respect the original dimensions of the systems, whereas they can
increase if the state is taken out of its original space of restricted
dimension. Let's take as an example the four-partite concurrence
$|\bra{\psi^{\ast}}\sigma_{y}^{\otimes 4}\ket{\psi}|$
\cite{verstaete2003,rigolin2005,wong2001},
which is a straightforward generalization of the original concurrence
$|\bra{\psi^{\ast}}\sigma_{y}\otimes\sigma_{y}\ket{\psi}|$ \cite{wootters1998}
with the complex conjugation performed in the standard separable basis. Its
definition restricts this quantity to system of four qubits only. And within
this restriction it is known to be a proper monotone \cite{verstaete2003},
since it is invariant under stochastic local operations assisted by classical
communication -- called SLOCC.

Whereas this definition of a four-partite concurrence applies only to qubit
systems, it can be generalized to systems of arbitrary dimensions via
Eq.~(\ref{eq:concurrence}) and the operator
\begin{equation}
\mathcal{A}=P_- \otimes P_- \otimes P_- \otimes P_-\ .
\end{equation}
Here, we drop the normalization, since our convention differs from those of
\cite{verstaete2003,rigolin2005,wong2001},
and is immaterial for the subsequent analysis.

Thus, one can check monotonicity with respect to general LOCC operations with
the criterion of Eq.~(\ref{theorem}). For the coefficients $\alpha_\mathbf{V}$
one obtains that $\alpha_{\mathbf{V}}=-1$ if $\mathbf{V}$ is a one-element or a
three-element subset, and $\alpha_\mathbf{V}=1$ for two-element subsets. As
already indicated by numerical results presented at the end of
Sec.~\ref{sec:monconcurrence}, for this values of $\alpha_\mathbf{V}$ the
concurrence is {\em not} an entanglement monotone.

Indeed, this can be seen analytically. Let's choose the exemplary states
$\ket{\psi}=\ket{\Phi^+}\otimes \ket{\Phi^+}$, and
$\ket{\phi}=\ket{\Phi^-}\otimes\ket{\Phi^-}$. They violate inequality (\ref{eq:condition1}),
since $C(\ket{\psi})=C(\ket{\phi})=1/2$, while
$C(\ket{\psi} \otimes \ket{\eta_{1}}/\sqrt{2}+ \ket{\phi}\otimes\ket{\eta_{2}}/\sqrt{2})=1/2\sqrt{2}$.

\section{Summary}
\label{sec:summary}
We investigated the monotonicity properties of generalized, concurrence-like
quantities which are useful for quantifying the entanglement of multipartite systems. The
concurrences, defined for pure states in terms of projection operators and
extended to mixed states with the help of the convex-roof procedure, posses a
range of attractive features:
They allow for an effective lower bound estimation, what leads to a discrimination between separable and entangled
states \cite{mintert2004a,mintert2005}.
They are useful for investigations of entanglement dynamics under decoherence,
and especially of the robustness of various
types of entangled states \cite{carvalho2005,mintert2005a}.
Moreover, they can
be used to implement experimentally effective estimations of entanglement
\cite{walborn06,aolita2006,mintert2006}.

The constructed generalized concurrences depend on continuous families of
parameters.
This gives the desired flexibility for adapting them to concrete
applications by choosing particular values of the parameters. On the other hand,
not all choices are allowed if we demand the concurrences to become genuine
entanglement measures. As such they should be non-increasing under LOCC.
In the present paper we adapted general conditions of monotonicity under LOCC which were
formulated in \cite{vidal2000}, and have been simplified in \cite{horodecki2004}.
We obtained a necessary and sufficient condition for
monotonicity Eq.~(\ref{eq:condition1}), valid for all measures defined via the
convex-roof construction (\ref{eq:cauchy}), which is one of the canonical ways for
defining mixed-state entanglement measures \cite{vidal2000}.
Based on this, we were able to give a simple, sufficient condition for the monotonicity
of the generalized concurrences (\ref{theorem}), which is also a necessary one in
two and three-partite cases.

The constructed monotonous measures of multipartite entanglement, being for
pure states always bilinear in the components of the state, cannot completely characterize
all classes of entangled states equivalent under LOCC.
Notwithstanding, this leaves their crucial merits -- the relative ease of a quantitative evaluation or
estimation, as well as the direct experimental accessibility -- unaffected.

\begin{acknowledgments}
We are indebted to Marco Piani for valuable discussions.
Financial support by VolkswagenStiftung (under the project ``Entanglement
measures and the influence of noise'') is gratefully acknowledged.
F.M. thanks the Alexander von Humboldt foundation for support within the Feodor Lynen program.
RDD acknowledges support from Foundation for
Polish Science and from the Polish Ministry of Science and Higher Education
under grant No 1~P03B~129~30.
\end{acknowledgments}


\end{document}